\documentclass[12pt]{article}

\usepackage{graphics, amssymb, theorem, latexsym, bbm, amsmath, subcaption, graphicx}
\usepackage[utf8]{inputenc}
\usepackage{natbib}
\UseRawInputEncoding
\begin{document}
\newcommand{\ncm}{\newcommand}
\ncm{\bfm}[1]{\mbox{\boldmath $#1$}}
\ncm{\sbfm}[1]{\mbox{\scriptsize\boldmath $#1$}}
\ncm{\scr}[1]{\mbox{\scriptsize #1}}
\ncm{\scrmath}[1]{\mbox{\scriptsize $#1$}}
\ncm{\bfmscr}[1]{\mbox{\scriptsize{\boldmath $#1$}}}

\ncm{\R}{{\mathbb{R}}}
\ncm{\Z}{{\mathbb{Z}}}
\ncm{\T}{{\mathbb{T}}}
\ncm{\Smath}{{\mathbb{S}}}
\ncm{\N}{{\mathbb{N}}}
\ncm{\C}{{\mathbb{C}}}
\ncm{\A}{{\mathbb{A}}}
\ncm{\amath}{\bfm{a}}
\ncm{\V}{{\mathbb{V}}}
\ncm{\Hap}{{\mathbb{H}}}
\ncm{\MMD}{{\mathbb{MD}}}

\ncm{\cA}{{\cal A}}
\ncm{\cB}{{\cal B}}
\ncm{\cC}{{\cal C}}
\ncm{\calF}{{\cal F}}
\ncm{\cD}{{\cal D}}
\ncm{\cG}{{\cal G}}
\ncm{\cL}{{\cal L}}
\ncm{\cN}{{\cal N}}
\ncm{\cI}{{\cal I}}
\ncm{\cJ}{{\cal J}}
\ncm{\cH}{{\cal H}}
\ncm{\cV}{{\cal V}}
\ncm{\cW}{{\cal W}}
\ncm{\cT}{{\cal T}}
\ncm{\cX}{{\cal X}}
\ncm{\cQ}{{\cal Q}}
\ncm{\cR}{{\cal R}}
\ncm{\cS}{{\cal S}}
\ncm{\cM}{{\cal M}}
\ncm{\cU}{{\cal U}}
\ncm{\cP}{{\cal P}}
\ncm{\cZ}{{\cal Z}}
\ncm{\cO}{{\cal O}}
\ncm{\cPzer}{{\cal P}_0}
\ncm{\cPone}{{\cal P}_1}
\ncm{\cPk}{{\cP_{\mbox{\scr{known}}}}}
\ncm{\cF}{{\cal F}}
\ncm{\cE}{{\cal E}}
\ncm{\cMD}{{\cal MD}}
\ncm{\tcV}{\tilde{\cal V}}
\ncm{\cCobs}{{\cal C}_{\scr{obs}}}

\ncm{\Om}{\Omega}
\ncm{\om}{\omega}
\ncm{\va}{\varepsilon}
\ncm{\vam}{\varepsilon_{\scr{max}}}
\ncm{\de}{\delta}
\ncm{\De}{\Delta}
\ncm{\ga}{\gamma}
\ncm{\Ga}{\Gamma}
\ncm{\la}{\lambda}
\ncm{\ka}{\kappa}
\ncm{\si}{\sigma}
\ncm{\Si}{\Sigma}
\ncm{\La}{\Lambda}
\ncm{\eps}{\epsilon}

\ncm{\bY}{\bfm{Y}}
\ncm{\bA}{\bfm{A}}
\ncm{\bB}{\bfm{B}}
\ncm{\bC}{\bfm{C}}
\ncm{\bD}{\bfm{D}}
\ncm{\bF}{\bfm{F}}
\ncm{\bI}{\bfm{I}}
\ncm{\bZ}{\bfm{Z}}
\ncm{\bG}{\bfm{G}}
\ncm{\bH}{\bfm{H}}
\ncm{\bL}{\bfm{L}}
\ncm{\bP}{\bfm{P}}
\ncm{\bQ}{\bfm{Q}}
\ncm{\bR}{\bfm{R}}
\ncm{\bS}{\bfm{S}}
\ncm{\bT}{\bfm{T}}
\ncm{\bU}{\bfm{U}}
\ncm{\bV}{\bfm{V}}
\ncm{\bM}{\bfm{M}}
\ncm{\bN}{\bfm{N}}
\ncm{\bW}{\bfm{W}}
\ncm{\bX}{\bfm{X}}
\ncm{\bu}{\bfm{u}}
\ncm{\bv}{\bfm{v}}
\ncm{\bw}{\bfm{w}}
\ncm{\bwpr}{\bfm{w}^\prime}
\ncm{\bhp}{\bfm{h}^\prime}
\ncm{\bc}{\bfm{c}}
\ncm{\bd}{\bfm{d}}
\ncm{\bm}{\bfm{m}}
\ncm{\bh}{\bfm{h}}
\ncm{\bn}{\bfm{n}}
\ncm{\bb}{\bfm{b}}
\ncm{\bg}{\bfm{g}}
\ncm{\be}{\bfm{e}}
\ncm{\bl}{\bfm{l}}
\ncm{\bp}{\bfm{p}}
\ncm{\bq}{\bfm{q}}
\ncm{\br}{\bfm{r}}
\ncm{\bs}{\bfm{s}}
\ncm{\bx}{\bfm{x}}
\ncm{\by}{\bfm{y}}
\ncm{\bz}{\bfm{z}}
\ncm{\balp}{\bfm{\alpha}}
\ncm{\bbe}{\bfm{\beta}}
\ncm{\bxi}{\bfm{\xi}}
\ncm{\bth}{\bfm{\theta}}
\ncm{\bom}{\bfm{\om}}
\ncm{\bmu}{\bfm{\mu}}
\ncm{\bde}{\bfm{\de}}
\ncm{\bva}{\bfm{\va}}
\ncm{\beps}{\bfm{\eps}}
\ncm{\bga}{\bfm{\ga}}
\ncm{\bka}{\bfm{\ka}}
\ncm{\bla}{\bfm{\la}}
\ncm{\bpi}{\bfm{\pi}}
\ncm{\bpsi}{\bfm{\psi}}
\ncm{\brho}{\bfm{\rho}}
\ncm{\boldeta}{\bfm{\eta}}
\ncm{\bphi}{\bfm{\phi}}
\ncm{\bLa}{\bfm{\Lambda}}
\ncm{\bPi}{\bfm{\Pi}}
\ncm{\bSi}{\bfm{\Si}}
\ncm{\bone}{\bfm{1}}

\ncm{\sbb}{\sbfm{b}}
\ncm{\sbc}{\sbfm{c}}
\ncm{\sbd}{\sbfm{d}}
\ncm{\sbC}{\sbfm{C}}
\ncm{\sbM}{\sbfm{M}}
\ncm{\sbX}{\sbfm{X}}
\ncm{\sbw}{\sbfm{w}}
\ncm{\sbx}{\sbfm{x}}
\ncm{\sby}{\sbfm{y}}
\ncm{\subv}{\scrmath{\ubv}}
\ncm{\subw}{\scrmath{\ubw}}
\ncm{\subx}{\scrmath{\ubx}}
\ncm{\suby}{\scrmath{\uby}}
\ncm{\subX}{\scrmath{\ubX}}

\ncm{\hbe}{\hat{\beta}}
\ncm{\heta}{\hat{\eta}}
\ncm{\hth}{\hat{\theta}}
\ncm{\hLa}{\hat{\Lambda}}
\ncm{\hSi}{\hat{\Sigma}}
\ncm{\hbth}{\hat{\bth}}
\ncm{\hh}{\hat{h}}
\ncm{\hs}{\hat{s}}
\ncm{\hB}{\hat{B}}
\ncm{\hD}{\hat{D}}
\ncm{\hF}{\hat{F}}
\ncm{\hI}{\hat{I}}
\ncm{\hN}{\hat{N}}
\ncm{\hP}{\hat{P}}
\ncm{\hQ}{\hat{Q}}
\ncm{\htau}{\hat{\tau}}
\ncm{\hga}{\hat{\gamma}}
\ncm{\hla}{\hat{\lambda}}
\ncm{\hmu}{\hat{\mu}}
\ncm{\hpi}{\hat{\pi}}
\ncm{\hpsi}{\hat{\psi}}
\ncm{\hbB}{\hat{\bB}}
\ncm{\hbbe}{\hat{\bbe}}
\ncm{\hbga}{\hat{\bga}}
\ncm{\hbpsi}{\hat{\bpsi}}

\ncm{\mast}{m^\ast}
\ncm{\cast}{c^\ast}
\ncm{\fast}{f^\ast}
\ncm{\siast}{\si^\ast}
\ncm{\psiast}{\psi^\ast}
\ncm{\tsiast}{\tilde{\si}^\ast}
\ncm{\alfast}{\alpha^\ast}
\ncm{\tkaast}{\tilde{\kappa}^\ast}

\ncm{\ap}{a^\prime}
\ncm{\hp}{h^\prime}
\ncm{\ip}{i^\prime}
\ncm{\jp}{j^\prime}
\ncm{\kp}{k^\prime}
\ncm{\np}{n^\prime}
\ncm{\npr}{n^\prime}
\ncm{\qp}{q^\prime}
\ncm{\spr}{s^\prime}
\ncm{\up}{u^\prime}
\ncm{\vp}{v^\prime}
\ncm{\wpr}{w^\prime}
\ncm{\xp}{x^\prime}
\ncm{\yp}{y^\prime}
\ncm{\zp}{z^\prime}
\ncm{\Cp}{C^\prime}
\ncm{\Gp}{G^\prime}
\ncm{\Ip}{I^\prime}
\ncm{\Mp}{M^\prime}
\ncm{\Np}{N^\prime}
\ncm{\Npr}{N^\prime}
\ncm{\Tp}{T^\prime}
\ncm{\gap}{\ga^\prime}
\ncm{\phpr}{\phi^\prime}

\ncm{\wbis}{w^{\prime\prime}}

\ncm{\tih}{\tilde{h}}
\ncm{\tZ}{\tilde{Z}}
\ncm{\tA}{\tilde{A}}
\ncm{\tD}{\tilde{D}}
\ncm{\tF}{\tilde{F}}
\ncm{\tI}{\tilde{I}}
\ncm{\tN}{\tilde{N}}
\ncm{\tQ}{\tilde{Q}}
\ncm{\tY}{\tilde{Y}}
\ncm{\tmu}{\tilde{\mu}}
\ncm{\tOm}{\tilde{\Omega}}
\ncm{\tnu}{\tilde{\nu}}
\ncm{\tsi}{\tilde{\sigma}}
\ncm{\tal}{\tilde{\alpha}}
\ncm{\tbeta}{\tilde{\beta}}
\ncm{\tde}{\tilde{\delta}}
\ncm{\txi}{\tilde{\xi}}
\ncm{\tmathV}{\tilde{\V}}
\ncm{\tV}{\tilde{V}}
\ncm{\tr}{\tilde{r}}
\ncm{\tu}{\tilde{u}}
\ncm{\tw}{\tilde{w}}
\ncm{\twpr}{\tilde{w}^\prime}
\ncm{\tb}{\tilde{b}}
\ncm{\td}{\tilde{d}}
\ncm{\tp}{\tilde{p}}
\ncm{\tf}{\tilde{f}}
\ncm{\tn}{\tilde{n}}
\ncm{\tS}{\tilde{S}}
\ncm{\tL}{\tilde{L}}
\ncm{\tl}{\tilde{l}}
\ncm{\tP}{\tilde{P}}
\ncm{\tSmath}{\tilde{\mathbb{S}}}
\ncm{\tT}{\tilde{T}}
\ncm{\tK}{\tilde{K}}
\ncm{\tka}{\tilde{\ka}}
\ncm{\tom}{\tilde{\om}}
\ncm{\tva}{\tilde{\va}}
\ncm{\tla}{\tilde{\la}}
\ncm{\tpi}{\tilde{\pi}}
\ncm{\trho}{\tilde{\rho}}
\ncm{\tbom}{\tilde{\bfm{\om}}}
\ncm{\tbxi}{\tilde{\bfm{\xi}}}
\ncm{\tbrho}{\tilde{\bfm{\rho}}}
\ncm{\tbg}{\tilde{\bg}}
\ncm{\tbb}{\tilde{\bb}}
\ncm{\tbr}{\tilde{\br}}
\ncm{\tbf}{\tilde{\bfm{f}}}
\ncm{\tbD}{\tilde{\bfm{D}}}
\ncm{\tbH}{\tilde{\bfm{H}}}
\ncm{\tbone}{\tilde{\bfm{1}}}
\ncm{\tbe}{\tilde{\bfm{e}}}
\ncm{\tbbe}{\tilde{\bfm{\beta}}}

\ncm{\bae}{\bar{e}}
\ncm{\baf}{\bar{f}}
\ncm{\bah}{\bar{h}}
\ncm{\bal}{\bar{l}}
\ncm{\bam}{\bar{m}}
\ncm{\ban}{\bar{n}}
\ncm{\bap}{\bar{p}}
\ncm{\bav}{\bar{v}}
\ncm{\baw}{\bar{w}}
\ncm{\baF}{\bar{F}}
\ncm{\baZ}{\bar{Z}}
\ncm{\baY}{\bar{Y}}
\ncm{\baS}{\bar{S}}
\ncm{\baH}{\bar{H}}
\ncm{\baA}{\bar{A}}
\ncm{\baD}{\bar{D}}
\ncm{\baC}{\bar{C}}
\ncm{\baN}{\bar{N}}
\ncm{\baQ}{\bar{Q}}
\ncm{\baW}{\bar{W}}
\ncm{\bacW}{\bar{\cW}}
\ncm{\bacV}{\bar{\cV}}
\ncm{\bacR}{\bar{{\cal R}}}
\ncm{\bacP}{\bar{{\cal P}}}
\ncm{\babe}{\bar{\beta}}
\ncm{\baka}{\bar{\kappa}}
\ncm{\bamu}{\bar{\mu}}
\ncm{\banu}{\bar{\nu}}
\ncm{\bade}{\bar{\de}}
\ncm{\bala}{\bar{\la}}
\ncm{\baga}{\bar{\ga}}
\ncm{\barho}{\bar{\rho}}
\ncm{\babf}{\bar{\bfm{f}}}
\ncm{\babD}{\bar{\bfm{D}}}
\ncm{\babA}{\bar{\bfm{A}}}
\ncm{\babQ}{\bar{\bfm{Q}}}
\ncm{\babW}{\bar{\bfm{W}}}
\ncm{\babh}{\bar{\bfm{h}}}
\ncm{\babr}{\bar{\bfm{r}}}
\ncm{\babde}{\bar{\bfm{\de}}}
\ncm{\babrho}{\bar{\bfm{\rho}}}
\ncm{\babone}{\bar{\bfm{1}}}

\ncm{\chnu}{\check{\nu}}

\ncm{\uC}{\underline{C}}
\ncm{\ucX}{\underline{\cX}}
\ncm{\ubx}{\underline{\bx}}
\ncm{\ubv}{\underline{\bv}}
\ncm{\ubw}{\underline{\bw}}
\ncm{\ubX}{\underline{\bX}}
\ncm{\uby}{\underline{\by}}
\ncm{\ubY}{\underline{\bY}}

\ncm{\Lin}{\, \stackrel{\cal L} \in}
\ncm{\Leq}{\, \stackrel{\cal L} =}
\ncm{\Lto}{\, \stackrel{\cal L} \longrightarrow}
\ncm{\pto}{\, \stackrel{p} \longrightarrow}
\ncm{\asto}{\, \stackrel{\rm a.s.} \longrightarrow}
\ncm{\Cov}{\mbox{Cov}}
\ncm{\Var}{\mbox{Var}}
\ncm{\sameord}{\stackrel{\cup}{{\scriptstyle \cap}}}

\ncm{\ith}{i^{\scr{th}}}
\ncm{\jth}{j^{\scr{th}}}
\ncm{\kth}{k^{\scr{th}}}
\ncm{\lth}{l^{\scr{th}}}
\ncm{\Bin}{\mbox{Bin}}
\ncm{\Exp}{\mbox{Exp}}
\ncm{\Hyp}{\mbox{Hyp}}
\ncm{\mm}{\mbox{mm}}
\ncm{\base}{\scr{bl}}
\ncm{\PD}{\mbox{PD}}
\ncm{\sgn}{\mbox{sgn}}
\ncm{\Ctot}{\bar{C}}
\ncm{\Ctottiny}{C_{\mbox{\tiny tot}}}
\ncm{\bzero}{\bfm{0}}
\ncm{\fappr}{\hat{f}}
\ncm{\bappr}{\hat{b}}
\ncm{\laappr}{\hat{\la}}
\ncm{\muappr}{\hat{\mu}}
\ncm{\pappr}{\hat{p}}
\ncm{\piappr}{\hat{\pi}}
\ncm{\kaappr}{\hat{\ka}}
\ncm{\Siappr}{\hat{\Si}}
\ncm{\bSiappr}{\hat{\bSi}}
\ncm{\demax}{\de_{\scr{max}}}
\ncm{\mumin}{\mu_{\scr{min}}}
\ncm{\hmumin}{\hmu_{\scr{min}}}
\ncm{\Ias}{I_{\scr{as}}}
\ncm{\Ibas}{I_B}
\ncm{\cBall}{\cB_{\scr{all}}}
\ncm{\Inonas}{I_{\scr{nas}}}
\ncm{\Ilong}{I_{\scr{long}}}
\ncm{\Ishort}{I_{\scr{short}}}

\ncm{\beq}{\begin{equation}}
\ncm{\eeq}{\end{equation}}
\ncm{\beqr}{\begin{eqnarray}}
\ncm{\eeqr}{\end{eqnarray}}
\ncm{\beqrn}{\begin{eqnarray*}}
\ncm{\eeqrn}{\end{eqnarray*}}
\ncm\rthm[1]{\ref{#1}}
\ncm\re[1]{(\ref{#1})}
\ncm{\slut}{
  {\unskip\nobreak\hfill\penalty100\hskip1em\vadjust{}\nobreak
  \hfill\mbox{$\Box$}\parfillskip=0pt\finalhyphendemerits=0}}

\parindent=0mm
\newcommand*\samethanks[1][\value{footnote}]{\footnotemark[#1]}

\newcommand{\MAP}{\hat{\theta}^{\text{(MAP)}}}
\newcommand{\Bayes}{\hat{\theta}^{\text{(Bayes)}}}
\ncm{\cDnew}{{\cal D}^{\text{new}}}
\ncm{\cDobs}{{\cal D}^{\text{obs}}}
\ncm{\cDsim}{{\cal D}^{\text{sim}}}
\ncm{\Ipr}{I^\prime}
\ncm{\hatom}{\hat{\omega}}
\ncm{\Prob}{\mathbb{P}}
\ncm{\E}{\mathbb{E}}
\ncm{\I}{\mathbbm{1}}
\newcommand{\mY}{\mathbf{Y}}
\newcommand{\mX}{\mathbf{X}}
\newcommand{\mB}{\mathbf{B}}
\newcommand{\mE}{\mathbf{E}}
\newcommand{\mZ}{\mathbf{Z}}
\newcommand{\mU}{\mathbf{U}}
\newcommand{\mC}{\mathbf{C}}
\newcommand{\mD}{\mathbf{D}}
\newcommand{\mV}{\mathbf{V}}
\newcommand{\mS}{\mathbf{S}}
\newcommand{\mI}{\mathbf{I}}
\newcommand{\bbeta}{\boldsymbol{\beta}}
\newcommand{\Sigmajuv}{\bSigma^\texttt{juv}}
\newcommand{\Sigmaad}{\bSigma^\texttt{ad}}
\newcommand{\sfI}{\mathsf{I}}
\newcommand{\classI}{\hat{\mathrm{I}}}
\newcommand{\rmI}{\mathrm{I}}
\newcommand{\sfN}{\mathsf{N}}
\newcommand{\sfS}{\mathsf{S}}
\newcommand{\sfK}{\mathsf{K}}
\newcommand{\Rtot}{R^\text{tot}}
\newcommand{\rmd}{\,\mathrm{d}}
\newcommand{\bSigma}{\mathbf{\Sigma}}
\newcommand{\wl}{\textit{wing length }}
\newcommand{\nl}{\textit{notch length }}
\newcommand{\Po}{\text{Po}}
\newcommand{\Geo}{\text{Geo}}
\newcommand{\Be}{\text{Be}}
\newcommand{\Wei}{\text{Wei}}
\newcommand{\U}{\text{U}}
\newcommand{\D}{\mathbf{D}}
\newcommand{\B}{\mathbf{B}}
\newcommand{\J}{\mathbf{J}}
\newcommand{\bt}{\mathbf{t}}
\newcommand{\Sigmat}{\mathbf{\Sigma}}
\newcommand{\Lambdat}{\mathbf{\Lambda}}
\newcommand{\lb}{\left\{}
\newcommand{\rb}{\right\}}
\newcommand{\lh}{\left[}
\newcommand{\rh}{\right]}
\newcommand{\lp}{\left(}
\newcommand{\rp}{\right)}
\newcommand{\Laplace}{\text{L}}
\newcommand{\convp}{\overset{p}{\longrightarrow}}
\newcommand{\convr}{\overset{r}{\longrightarrow}}
\newcommand{\convd}{\overset{d}{\longrightarrow}}
\newcommand{\tntoinf}{\text{när $n\to\infty$}}
\newcommand{\ntoinf}{$n\to\infty$}
\newcommand{\sumiton}{\sum_{i=1}^n}
\newcommand{\Tloc}{T_{\text{loc}}}
\newcommand{\Tscale}{T_{\text{scale}}}
\newcommand{\Tskew}{T_{\text{skew}}}
\newcommand{\Tkurt}{T_{\text{kurt}}}
\newcommand{\logit}{\text{logit}}

\title{A Comparison Between Quantile Regression and Linear Regression on Empirical Quantiles for Phenological Analysis in Migratory Response to Climate Change \\
\large Running title: Quantile regression for phenological analyses}

\author{M{\aa}ns Karlsson\thanks{Department of Mathematics, Stockholm University, 106 91 Stockholm, Sweden, mansk@math.su.se}
\\
Ola H\"{o}ssjer\thanks{Department of Mathematics, Stockholm University, 106 91 Stockholm, Sweden}}
\maketitle
\clearpage

\begin{abstract}
It is well established that migratory birds in general have advanced their arrival times in spring, and in this paper we investigate potential ways of enhancing the level of detail in future phenological analyses. We perform single as well as multiple species analyses, using linear models on empirical quantiles, non-parametric quantile regression and likelihood-based parametric quantile regression with asymmetric Laplace distributed error terms. We conclude that non-parametric quantile regression appears most suited for single as well as multiple species analyses.

\end{abstract}

\clearpage

\section{Introduction}
Analysis of ``the timing of seasonal activities of animals and plants'' is termed \textit{phenology} \citep{walther2002ecological}, and it is a well studied topic. For instance, in the meta analysis of \citet{usui2017temporal}, 73 different studies analysing bird phenology by means of linear regression were used to investigate the phylogenetic signal in the response to climate change. The problem can be approached in various ways, such as investigating if there is a connection between migration timing change and variations of some climate index \citep{jonzen2006rapid}, or if there is a trend in the location shift of the arrival distribution \citep{lehikoinen2019phenology}. These two approaches were used in the metastudy of \citet{usui2017temporal}.
\par\medskip
Phenological studies use data collected at one or several locations and may cover one or multiple species. When birds are studied, the data used is often collected in a systematic manner at bird observatories, since these organisations tend to have the long running series of data needed for phenological analyses (see \citet{knudsen2007characterizing} for an in depth study). In more recent years, citizen science data has been used as well \citep{mayer2010phenology}. The sample unit of the data is usually individual birds, and these might be trapped and ringed or just observed and registered without trapping, cf. \citet{lehikoinen2019phenology} for a study where both of these data collection methods were combined.
\par\medskip
Associated with each bird is a varying number of covariates. Every observation is associated with an arrival or departure date and almost always a species information, which can be used as a block factor in multiple species analyses. Ringed birds are also often registered to the hour of the day, and, if possible, the age and sex of a bird is determined as well. Other biometrics are sometimes collected, at least if the bird is captured.
\par\medskip
The main interest in the studies of \citet{jonzen2006rapid} and \citet{lehikoinen2019phenology}, as well as in this study, concerns changes over time in the distribution of arrival times. We will use {\it year} as the covariate of primary interest, as the interpretation of the associated coefficient will tell us what change has occured over time. Other covariates may also be used, such as the age and sex of birds or other biometrics \citep{aharon2021limited}. The empirical arrival distribution of any particular year can be visualized as a step function with variable step sizes. The step sizes depend on the regional and local weather at the study site, and a study of the weather effects for fall migration of European Robins has been conducted by \citet{karlsson2014modelling}. In this study we do not include weather covariates though.
\par\medskip
\citet{knudsen2007characterizing} attempted two ways of smoothing these yearly arrival distributions, before fitting a model to data, and they concluded that methodological advances were needed. In this paper we will not perform any smoothing of this kind before fitting models to data.
\par\medskip
The purpose of this article is to compare linear models for empirical quantiles with linear quantile models without \citep{koenker2005quantile} or with \citep{geraci2014linear} random effects. The latter model, which essentially is a mixed effects model for quantile regression, has recently been employed for phenological analyses \citep{aharon2021limited}. The remainder of our paper is structured as follows. In Section \ref{statapp}, fixed effect models are first introduced for single species data and then fixed and mixed effects models are presented for multiple species data. Section \ref{data} contains a presentation of the data we will analyse, Section \ref{results} provides a summary of the results of the analysis, and finally in Section \ref{discussion} we discuss potential improvements and extensions. Some mathematical details are gathered in the supplementary material.

\section{Statistical approaches} \label{statapp}
This section contains an overview of the statistical methods used in this paper for phenological analysis. We will cover the empirical quantiles linear models used in \citet{jonzen2006rapid} and \citet{lehikoinen2019phenology}, nonparametric quantile regression \citep{koenker2005quantile} and a likelihood-based quantile regression approach using the asymmetric Laplace distribution \citep{geraci2014linear}. These methods are described for singular species as well as for multiple species analyses.

\newcommand{\Yeq}{Y^{\text{eq}}}
\newcommand{\Xeq}{X^{\text{eq}}}
\newcommand{\betaeq}{\beta^{\text{eq}}}
\newcommand{\sigmaeq}{\sigma_{\text{eq}}}
\newcommand{\varepsiloneq}{\varepsilon^{\text{eq}}}
\newcommand{\betaqr}{\beta^{\text{qr}}}
\newcommand{\varepsilonqr}{\varepsilon^{\text{qr}}}
\newcommand{\betalqm}{\beta^{\text{lqm}}}
\newcommand{\sigmalqm}{\sigma_{\text{lqm}}}

\subsection{Single species models} \label{sec:sspm}
Throughout this paper, the response variable will be \textit{julian day}. For single species models, $y_j$ will be the julian arrival day of individual $j \in \{1,\ldots,n\}$. With each observation comes a covariate vector $x_j = \left(1, t_j, x_{j1}, \ldots, x_{jp}\right)$, consisting of an intercept term $1$, the year $t_j$ that individual $j$ was recorded, and $p$ additional covariates. Typical covariates are the binary \textit{age} (juvenile or adult) and \textit{sex} (female or male), but continuous covariates such as \textit{wing length} may also be used \citep{aharon2021limited}. We assume that $t_j$ takes values in ${\cal T} = \{1,\ldots,T\}$ for all $j$ and that $x_{j1},\ldots,x_{jp}$ take values in ${\cal X}_1, \ldots, {\cal X}_p$ respectively.

\subsubsection{Empirical quantile linear models} \label{sec:eqlm}
The empirical quantile (eq) model does not take the raw julian day $y_j$ as response, but rather an empirical quantile of the julian days of a subset of observations. Let ${\cal X} = {\cal X}_{1} \times \ldots \times {\cal X}_{p}$. For each pair $(x,t) \in {\cal X} \times {\cal T}$ we extract the set of observations
\begin{equation}
 {\cal Y}_{(x,t)} = \lb y_{j} : 1\left(t_{j} = t \land (x_{j1},\ldots,x_{jp}) = x\right) \rb,
\end{equation}
where $1(\cdot)$ is an indicator function that equals 1 if the condition within the brackets is true, and 0 otherwise. We also create the $T \cdot |{\cal X}| \times (2+p)$ matrix $\Xeq$ by stacking all vectors $(1, t, x)$ on top of each other. Let $\tau\in (0,1)$ be a quantile. For each of our sets ${\cal Y}_{(x,t)}$ we let $\hat{F}_{(x,t)}$ be the empirical distribution function formed by the elements of this set, and define the corresponding empirical quantile 
\begin{equation}
 \hat{Q}_{(x, t)}(\tau) = \inf_y\lb y \in {\cal Y}_{(x, t)} : \hat{F}_{(x,t)}(y) \geq \tau \rb,
\end{equation}
conditional on ${\cal Y}_{(x, t)} \neq \emptyset$, and stack these quantiles, for all $(x,t)\in {\cal T}\times {\cal X}$, into the vector $\Yeq(\tau)$. We then formulate the linear model
\begin{align} \label{eqlm}
 \Yeq(\tau) = \Xeq\betaeq(\tau) + \varepsiloneq(\tau)
\end{align}
where (omitting $\tau$) the column vector $\betaeq = (\betaeq_0, \betaeq_t, \betaeq_1, \ldots, \betaeq_p)^\top$ contains the intercept $\betaeq_0$, slope on \textit{year} $\betaeq_t$ and the $p$ covariate effects $\betaeq_1, \ldots, \betaeq_p$, whereas $\top$ refers to matrix transposition. The vector of error terms in $\varepsiloneq(\tau)$ is assumed to have a multivariate normal distribution $N\left(0,\sigmaeq^2(\tau) I_{T\cdot |{\cal X}|}\right)$, where $I_{T\cdot |{\cal X}|}$ is the identity matrix of rank $T\cdot |{\cal X}|$.
This approach gives each combination of covariate values and \textit{year} the same weight. However, if we want to give each recorded bird the same weight, some adjustements are needed. Omitting any dependence on $\tau$ in the notation, the log-likelihood for the model in \eqref{eqlm} is
\begin{equation} \label{lmll}
 l\lp \betaeq, \sigmaeq^2 \mid \Yeq, \Xeq\rp = \sum_{(x,t) \in {\cal T}\times{\cal X}} \log f \lp \hat{Q}_{(x,t)}(\tau) \mid (\Xeq\betaeq)_{(x,t)}, \sigmaeq^2 \rp
\end{equation}
where $f(\cdot \mid m,\sigma^2)$ is the Gaussian density with mean $m$ and variance $\sigma^2$. Let $w_{(x,t)}$ denote a weight given to each empirical quantile $\hat{Q}_{(x,t)}(\tau)$. If we let $w_{(x,t)} = |{\cal Y}_{(x,t)}|$, we will achieve equal weighting of each bird by reweighting the log-likelihood \eqref{lmll} as
\begin{equation} \label{lmllw}
 l_w\lp \betaeq, \sigmaeq^2 \mid \Yeq, \Xeq\rp = \sum_{(x,t) \in {\cal T}\times{\cal X}} w_{(x,t)} \log f \lp \hat{Q}_{(x,t)}(\tau) \mid (\Xeq\betaeq)_{(x,t)}, \sigmaeq^2 \rp.
\end{equation}
As is well known, the objective function \eqref{lmllw} is continous, concave and twice differentiable, and thus it is easily optimized in order to find the MLE of $\betaeq$ and $\sigmaeq^2$. This model fitting can then be repeated for any $\tau$ of interest.

\subsubsection{Non-parametric quantile regression} \label{sec:qr}
For a deeper treatment of quantile regression (qr), we refer the reader to the standard litterature of \citet{koenker2005quantile}, which is the primary source for the material that is presented in this section.
\par\medskip
Instead of a linear predictor for the mean of the response distribution, a quantile regression model has a linear predictor for a conditional quantile of the response distribution. As before, let $y_j$ be the observed response for individual $j$, and stack all the covariate vectors $x_j = (1,t_j,x_{j1},\ldots,x_{jp})$ into the $n \times (p+2)$-matrix $X$. Let the conditional distribution function of the random variable $Y_j$ corresponding to its observed value $y_j$ be 
\begin{equation} \label{empdist}
 F_{Y_j\mid x_j}(y) = \Prob(Y_j \le y \mid x_j), \quad -\infty < y < \infty.
\end{equation}
For each $0<\tau <1$ the inverse
\begin{equation} \label{empquant}
 Q(\tau \mid x_j) = \inf \{y; F_{Y_j\mid x_j}(y) \ge \tau\}
\end{equation}
of (6) is the conditional quantile function, and we may construct models of the form
\begin{equation} \label{qrmod}
 Q\lp \tau \mid X \rp = X\betaqr(\tau) + \varepsilonqr(\tau)
\end{equation}
where (omitting $\tau$) $\betaqr = (\betaqr_0, \betaqr_t, \betaqr_1, \ldots, \betaqr_p)^\top$ is analogous to $\betaeq$. No parametric assumptions are made about the error terms $\varepsilonqr(\tau)$.
\par\medskip
The objective function in \eqref{qrmod} is the loss function
\begin{equation} \label{qrobj}
 \sum_{j=1}^n \rho_\tau(y_j - x_jb)
\end{equation}
where 
\begin{equation} \label{rho}
\rho_\tau(\nu) = \nu (\tau - 1(\nu < 0))
\end{equation}
\citep[p. 5]{koenker2005quantile}. We thus choose $\tau$ and estimate the regression parameters by finding
\begin{equation} \label{qrest}
 \betaqr(\tau) = \arg\min_{b\in\R^{2+p}} \sum_{j=1}^n \rho_\tau(y_j - x_jb).
\end{equation}
As opposed to \eqref{lmllw}, the objective function in \eqref{qrest} is not differentiable everywhere. In more detail, the derivative of the objective function has discontinuities along hyperplanes $\{b; \, y_j=x_jb\}$ for $j=1,\ldots,n$. Non-gradient based numerical estimation procedures are however available, and we use the Frisch-Newton interior point method \citep{portnoy1997gaussian} provided in the \textsf{R} package \texttt{quantreg} \citep{quantreg2021quantile}.
\par\medskip
As with the empirical quantile models, we may fit quantile regression models to any number of quantiles $\tau$. Since \eqref{qrest} applies directly to raw data $\{(x_j,y_j); \, j=1,\ldots,n\}$ on the sample unit level, no reweighting is needed to give each bird the same weight.

\subsubsection{Linear quantile models} \label{sec:lqm}
The third approach to modelling a single species is the linear quantile model (lqm). It is based on the assumption that the error terms in (\ref{qrmod}) are independent and identically distributed, and follow an asymmetric Laplace (AL) distribution. A random variable $\Gamma \sim AL(\mu, \sigma, \tau)$ if it has probability density function
\begin{equation} \label{pdflaplace}
 f_\gamma(\mu, \sigma, \tau) = \frac{\tau(1-\tau)}{\sigma}\exp\lb-\frac{1}{\sigma}\rho_\tau(\gamma - \mu)\rb,
\end{equation}
where $\mu \in (-\infty,\infty)$ is the location parameter, $\sigma > 0$ is the scale parameter and $\tau \in (0,1)$ is the asymmetry parameter \citep{hinkley1977estimation}. Note that
\begin{align}
 \Prob\lp\Gamma \leq \mu \rp &= \int_{-\infty}^\mu \frac{\tau(1-\tau)}{\sigma}\exp\lb-\frac{1}{\sigma}\rho_\tau(\gamma - \mu)\rb \rmd \gamma \\
 &= \frac{\tau(1-\tau)}{\sigma} \int_{-\infty}^\mu \exp\lb (\gamma - \mu)\frac{1-\tau}{\sigma}\rb \rmd \gamma \\
 &= \tau \lh \exp\lb (\gamma - \mu)\frac{1-\tau}{\sigma}\rb \rh_{\gamma = -\infty}^{\gamma=\mu} \\
 &= \tau
\end{align}
meaning that $\mu$ is the $\tau$:th quantile of the response variable distribution, when it is assumed to be asymmetric Laplacian. Therefore, inference about the location parameter $\mu$ will be equivalent to inference about the $\tau$:th quantile of the response variable distribution $F_Y$. For further details, see e.g. \citet{koenker1999goodness}.
\par\medskip
Assume for simplicity that the error terms in (\ref{qrmod}) have a scale parameter $\sigmalqm = 1$. Then the log-likelihood to maximize under the asymmetric Laplace assumption is
\begin{equation} \label{llal}
 l(\betalqm, \sigmalqm \mid \tau, Y, X) = n\log(\tau(1-\tau)) - \sum_{j=1}^n \rho_\tau(y_j - x_j\betalqm(\tau)).
\end{equation}
A gradient of the objective function \eqref{llal} is defined by \citet{bottai2015gradient}, and it is used in a gradient-based search algorithm in order to find an estimate of $\betalqm$. In a large sample simulation study of the \texttt{quantreg}-package, \citet{bottai2015gradient} found computational advantages of the gradient search method when compared with the abovementioned Frisch-Newton approach. They also noted that the bias (approximated as the difference between the estimated and true parameter values, average over many samples) was practically zero for both the gradient search based method and the Frisch-Newton method. They did not comment on the largest observed difference between true parameter value and parameter estimate, but the largest observed relative absolute difference in the parameter estimates of the two methods was 0.12.
\par\medskip
In conclusion, since minimizing the objective function in \eqref{qrest} with respect to the regression parameter is equivalent to maximizing the log likelihood in \eqref{llal}, the linear quantile model provides the same parameter estimates as the quantile regression model of the previous section, if the same numerical estimation algorithm is used for parameter optimization. In Section \ref{sec:single} we will investigate how the choice of optimization method affects parameteter estimates.   

\subsection{Multiple species models} \label{sec:mult}
To incorporate several species in a phenological analysis, we can either include \textit{species} as a categorical predictor, and add interaction effects between this and other covariates accordingly, or we can consider species a random block factor and make use of mixed effects models \citep{lehikoinen2019phenology, jonzen2006rapid}. We will consider the categorical predictor approach for nonparameteric quantile regression (qr), and the block factor approach for empirical quantile mixed models (meq) as well as for the Linear Quantile Mixed Model (lqmm) approach \citep{geraci2014linear}.
\par\medskip
In the case of the presence of a block factor (i.e. \textit{species}) in data, the response is $y_{ij}$ where $i = 1,\ldots,M$ denotes which species the observation belongs to, and $j=1,\ldots,n_i$ index the observations within species $i$. In total, we have $\sum_{i=1}^M n_i = N$ observations. Analogously the covariate vector associated with each observation is now $x_{ij} = \lp 1, t_{ij}, x_{ij1},\ldots,x_{ijp} \rp$. 
We introduce the vectors $z_{ij} = \lp 1, t_{ij}, z_{ij1},\ldots,z_{ijq} \rp$, for $i=1,\ldots,M$ and $j=1,\ldots,n_i$. These vectors $z_{ij}$ will be used below to introduce a vector $u$ with $q+2$ random effects, for the meq and the lqmm models. The first two components of this vector correspond to a random intercept and slope for each species. 

\subsubsection{Empirical quantile linear mixed effects models} \label{eqmm}

\newcommand{\Ymeq}{Y^{\text{meq}}}
\newcommand{\Xmeq}{X^{\text{meq}}}
\newcommand{\Zmeq}{Z^{\text{meq}}}
\newcommand{\betameq}{\beta^{\text{meq}}}
\newcommand{\umeq}{u^{\text{meq}}}
\newcommand{\sigmameq}{\sigma_{\text{meq}}}
\newcommand{\varepsilonmeq}{\varepsilon^{\text{meq}}}

With the introduction of the block factor \textit{species}, we repeat the process of computing empirical quantiles in Section 2.1.1 for each species $i\in {\cal M} = \{1,\ldots,M\}$ and set of years ${\cal T}_i$. In order to simplify our exposition it is assumed in this section that $q=0$ and $z_{ij}=(1,t_{ij})$ for all birds $(i,j)$. The multispecies empirical quantile (meq) model of species $i$ then takes the form 
\begin{equation} \label{eqlmm}
 \Ymeq_i(\tau) = \Xmeq_i\betameq + \Zmeq_i\umeq_i+ \varepsilonmeq_i(\tau)
\end{equation}
where the $\lvert{\cal T}_i\rvert \cdot \lvert{\cal X}_i\rvert \times (2+p)$ matrix $\Xmeq_i$ is created by stacking the vectors $(1,t_i,x_i)$, with $x_i\in {\cal X}_i$, where ${\cal X}_i$ is the set of values $(x_{ij1},\ldots,x_{ijp})$ can take, whereas $\Zmeq_i$ consists of the first two columns of $\Xmeq_i$. The random effects vector $\umeq_i$ has a two-dimensional Gaussian distribution with expected value $(0,0)^\top$ and covariance matrix $\Psi$. The likelihood contribution of species $i$ (cf. Section 2.1.1) is thus obtained by integrating over the unobserved random effects, i.e.
\begin{equation} \label{meqL}
 \begin{split}
 L_i\lp \Psi,\beta, \sigma \mid y\rp &= \\
 \prod_{(x, t) \in {\cal X}_i\times{\cal T}_i} &\int_{\mathbb{R}^q} f \lp \hat{Q}_{(x, t)}(\tau) \mid \left(\Xmeq_i \betameq + \Zmeq_iu_i\right)_{(x, t)}, \sigma \rp f(u_i\mid\Psi) \rmd u_i.
 \end{split}
\end{equation}
To weight each bird equally, we again follow Section 2.1.1 and create weights $w_{(i,t)}$ based on the distribution of observations across \textit{year} and other covariates. The reweighted log likelihood contribution of species $i$ is then
\begin{equation}
 l_i \lp \Psi, \beta, \sigma \mid y\rp = \sum_{(x, t) \in {\cal X}_i\times{\cal T}_i} w_{(x,t)} \log \lp \int_{\mathbb{R}^q} f \lp \hat{Q}_{(x, t)}(\tau) \mid \ldots \rp f(u_i\mid\Psi) \rmd u_i \rp,
\end{equation}
with the same integration of random effects as in \eqref{meqL}. By summing the reweighted log likelihood over all species we obtain the log likelihood
\begin{equation}
 l \lp \Psi, \beta, \sigma \mid y\rp = \sum_{i=1}^M l_i \lp \Psi, \beta, \sigma \mid y\rp
\end{equation}
of the full dataset. We will fit the linear mixed effects models using the \textsf{R} \citep{r2021} package \texttt{lme4} package \citep{bates2015fitting}. Details of the model specification and computational approaches are given in the \texttt{lme4}-package vignette 
of \citet{bates2014computational}.



\subsubsection{Linear quantile mixed models} \label{lqmm}
The framework called linear quantile mixed models (lqmm) was proposed by \citet{geraci2014linear} as an extension of the random intercept quantile regression model of \citet{geraci2007quantile}, and it is implemented in the \textsf{R}-package \texttt{lqmm} \citep{geraci2014lqmm}. The aim of lqmm is to provide an analogue of linear mixed effects models in the setting of quantile regression, exploiting the connection  between the asymmetric Laplace likelihood and the objective function \eqref{qrobj} of a nonparametric quantile regression model \citep{koenker1999goodness}.
\par\medskip
In an lqmm for quantile $\tau$, the linear predictor is
\begin{equation}
 \mu_{ij}(\tau) = x_{ij}\beta(\tau) + z_{ij}u_i
\end{equation}
and it is assumed that 
\begin{equation}
 y_{ij} \mid u_i \sim AL \lp \mu_{ij}(\tau), \sigma_\varepsilon(\tau), \tau \rp
\end{equation}
independently for all $i,j$, with $\varepsilon_{ij}(\tau) = y_{ij} - \mu_{ij}(\tau)$ for all $i$ and $j$. The random effects vector $u_i$ is independent of $\{\varepsilon_{ij}(\tau)\}_{j=1}^{n_i}$ for any $i \in \cal M$, and, for our models, it is assumed to be multivariate Gaussian with covariance matrix $\Psi(\tau)$. Omitting in the notation the dependence of parameters on $\tau$, the marginal likelihood 
\begin{equation}\label{lqmmL}
L\lp \beta, \sigma_\varepsilon, \Psi \mid \lb y_{ij}\rb \rp = \prod_{i=1}^M \int_{\mathbb{R}^{q+2}} \prod_{j=1}^{n_i} f(y_{ij} \mid \mu_{ij}, \sigma_\varepsilon, \tau) p\lp u_i \mid \Psi \rp \rmd u_i
\end{equation}
is obtained by integrating out the random effects of each species. In \citet{geraci2014linear} the integral of each term of (\ref{lqmmL}) is approximated numerically with a Gaussian quadrature procedure.
\par\medskip
\citet{geraci2014linear} prove that the likelihood \eqref{lqmmL} is log-concave. They also make use of the iterative gradient-based optimization algorithm of \citet{bottai2015gradient} in order to find the argmax of \eqref{lqmmL}. In this context they also point out that derivative free methods are also viable for maximizing \eqref{lqmmL}, namely the Nelder-Mead derivative free optimization \citep{nelder1965simplex} in the \texttt{optim} function in \textsf{R}. Convergence is, in theory, guaranteed by the log-concavity of the likelihood, but when maximizing a numerical approximation of \eqref{lqmm}, log-concavity is not guaranteed and parameter estimates may depend on starting values, as mentioned in \citet[p.14]{geraci2014lqmm}.

\subsubsection{Non-parameteric quantile regression with interactions}
Since nonparametric quantile regression (qr) does not include random effects, we will instead include species as a fixed effect and let it interact with \textit{year}. The purpose is to compare the quantile regression estimates with estimates obtained from the meq and lqmm models of Sections \ref{eqmm} and \ref{lqmm}. In particular, for the latter two approaches, we will study inference conditional on predictions of the random effects $u_i$, $i=1,\ldots,M$.
\par\medskip
Following Section 2.1.2, we let each species $i$ contribute with two covariates, a species specific intercept parameter $\beta_{0i}$ and a species specific slope parameter $\beta_{ti}$, which is interpreted as an interaction parameter between {\it species} and {\it year}. In order not to overparametrize, species $i=1$ will be included in the intercept $\beta_0$, and the overall slope on \text{year} $\beta_t$. The model, in the special case $p=0$ and $x_{ij}=(1,t_{ij})$, is thus 
\begin{equation} \label{qrmult}
 Q(\tau \mid x_{ij}) = \betaqr_0 + \betaqr_tt_{ij} + 1(i>1)\left(\betaqr_{0i} + \betaqr_{ti}t_{kj}\right) + \varepsilonqr_{ij},
\end{equation}
for $i=1,\ldots,n$, keeping dependence on $\tau$ only on the left-hand side of (\ref{qrmult}). Notice that the interpretation of $\betaqr_{0i}$ and $\betaqr_{ti}$ is the difference in intercept and slope of species $i>1$ compared to species $1$.

\subsection{Predicting $u$}
Inference in mixed models can be marginal, i.e. looking at the fixed effects estimates $\hat{\beta}$ only, or conditional, i.e. looking at the species specific estimates of quantities obtained by conditioning on the random effects $u = \lp u_1,\ldots,u_M \rp$. Conditional inference requires predictions $\hat{u}$ of the random effects, which we now present for the linear mixed model of Section \ref{eqmm} and the lqmm of Section \ref{lqmm}.

\subsubsection{Linear mixed models}

We use the \texttt{lmer} function of the \texttt{lme4}-package  \citep{bates2015fitting} to fit linear mixed effects models. The predicted random effects $\hat{u}$ are extracted with the function \texttt{ranef}, and the value obtained is the mode of the conditional density of the random effects, conditionally on the response and the estimates of the other model parameters \citep[sec. 5.1.4]{bates2014computational}.

\subsubsection{Linear quantile mixed models}
An adaption of the \texttt{ranef} function is provided in the \texttt{lqmm}-package, named \texttt{ranef.lqmm}. We will briefly describe the random effects output of this function and how we adapted the prediction of $u$ to be more computationally efficient for our models. The full details of our adaption are provided in a supplementary document.
\par\medskip
Given the lqmm of Section 2.2.2, we let $Z$ be the random effects design matrix, $X$ the fixed effects design matrix and $y$ the response vector. The function \texttt{ranef.lqmm} gives the estimated best linear predictor 
\begin{equation} \label{ueblp}
 \hat{u}(\tau) = \hat{\Psi}(\tau) Z^\top \hat{\Sigma}^{-1} \lb y - X\hat{\beta}(\tau) - \hat{\mathbb{E}}\lh \varepsilon(\tau) \rh \rb,
\end{equation}
of the random effects in an lqmm \citep[eq. 12]{geraci2014linear}, where $\hat{u}(\tau)$ is an $M \times (q+2)$ matrix of the estimated random effects and $\hat{\Sigma}$ is an estimate of
\begin{equation} \label{covmat}
\Sigma = \Cov(Y) = Z \Psi(\tau) Z^\top + \Cov(\varepsilon(\tau)),
\end{equation}
the $N\times N$ block-diagonal covariance matrix of the response vector $Y$, of which $y$ is an observation. Note that $\Cov(\varepsilon(\tau)) = \sigma_\varepsilon(\tau) I_N$, where $I_N$ is the identity matrix of rank $N$ and $\sigma_\varepsilon(\tau)$ is the scale parameter of the asymmetric Laplace distribution for quantile $\tau$. We thus need to invert $\hat{\Sigma}$ to compute $\hat{u}(\tau)$, and since \texttt{ranef.lqmm} tries to invert $\Sigma$ head-on by a call to \texttt{solve}, this is sometimes unfeasible for large data sets on consumer grade computers. 
Therefore, we adapted \eqref{ueblp}-\eqref{covmat} by inverting $\Sigma$ in blocks, and thereby made prediction of $u$ a lot faster. The full details of the adapted computational approach are given in a supplementary document. We tested our adapted matrix inversion against the \texttt{ranef.lqmm} output for a subset of data, and the two approaches produced the same results.

\section{Data} \label{data}
The phenological data used in this paper has been collected and kindly provided for use by the Falsterbo Bird observatory. This section gives an overview of the data collection method and how data was filtered. 


\subsection{Data collection}
The Falsterbo Bird Observatory (55$^{\circ}$23'01.6"N, 12$^{\circ}$49'00.9"E) uses a method for systematic trapping and ringing of birds described in \citet{roos1981ring}. In short, an effort is made most days within a fixed number of dates each year (called a season) to capture birds using mist nets at fixed locations. The number of mist nets vary depending on the weather, to ensure bird welfare, as the birds may be harmed under certain weather conditions. The nets are active for at least four hours a day, but some days they might be active for longer, and some days for a shorter time, depending on the weather.
\par\medskip
The varying effort to capture birds poses some challenges, that we will treat in a simple manner in this paper, since it is not of primary interest. If the effort is extended for more than four hours any given day, a larger portion of all the birds that could possibly be captured that day is expected to be captured. This may cause the resulting arrival distribution to be biased towards days with a larger effort. As a simple corrective measure, we only include birds captured in the first four hours of effort any given day. Ideally, one would however include data on all captured birds in phenological analyses and correct for the varying sampling effort.
\par\medskip
We will work under the assumption that the number of ringed birds a given day is proportional to the number of birds migrating past the study site the preceeding night. The assumption is not fully verified. The correlation between nocturnal migratory activity and the number of ringed birds during autumn migation in Falsterbo has been studied in \citet{zehnder2001ringing}. They found a positive correlation under the assumption of weather affecting the number of trapped birds in the same way as it effects the migratory activity. Since we work with spring migration, patterns and behavior might be different. The general perception from field work is that most days with weather such that the welfare of caught birds is risked, correlate negatively with the number of birds. This means that a smaller number of birds is expected to be captured these days. However, some days such weather conditions rather cause a high number of birds to congregate in Falsterbo.

\subsection{Data selection and overview}
In total, $151\,992$ birds of 89 species have been recorded during the spring seasons of 1980-2019. We filter these data to remove birds recorded after the first four hours of effort, and species with less than 150 recorded birds in total, which yields a data set of $137\,870$ birds distributed over 38 species. From these, we select 25 species that can be assumed to mostly consist of migratory individuals but with few locally breeding birds mixed in. This data set is used for all multiple species models of Section \ref{sec:mult}.
\par
\begin{figure}
 \includegraphics[width=\textwidth]{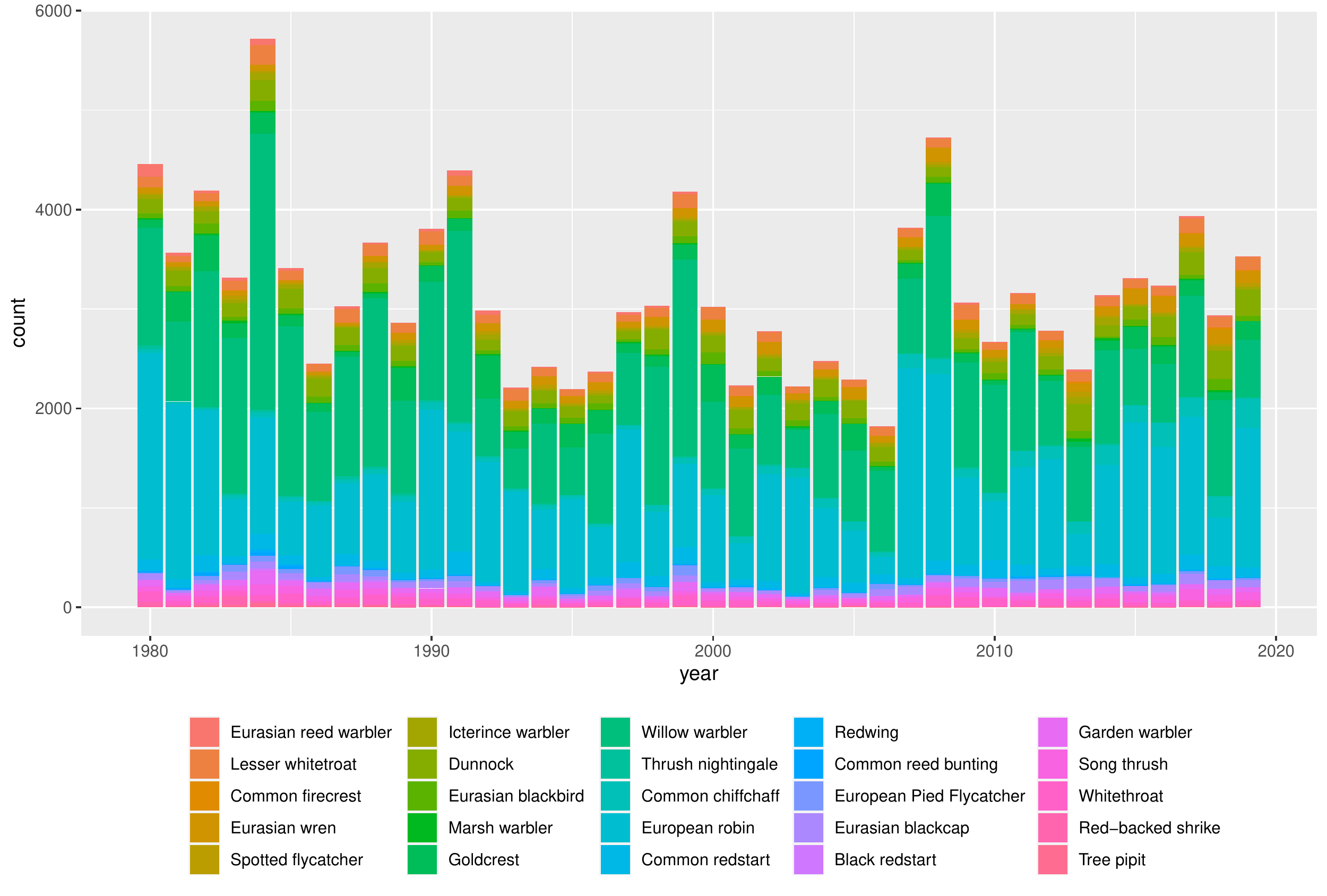}
 \caption{A color-coded barplot of the number of birds of each species each year. The plot is intended to illustrate the numerical dominance of European robin and Willow warbler in particular, and that year totals are quite consistent over time. Specific numbers are available in the supplementary material.}
 \label{balplot}
\end{figure}

\medskip
The data is very unbalanced with respect to species. In Figure \ref{balplot} illustrates the distribution of number of observations of each species, each year. There are two species that are very numerous, and the other species usually have much fewer birds each year. This could be an argument to rebalance data so that each species has the same weight in a mixed effects model. However, the fixed effects would not be interpretable then as the overall effects of birds in Falsterbo, but rather the effects when weighting up scarcely recorded species.
\par\medskip
\begin{figure}
 \includegraphics[width=\textwidth]{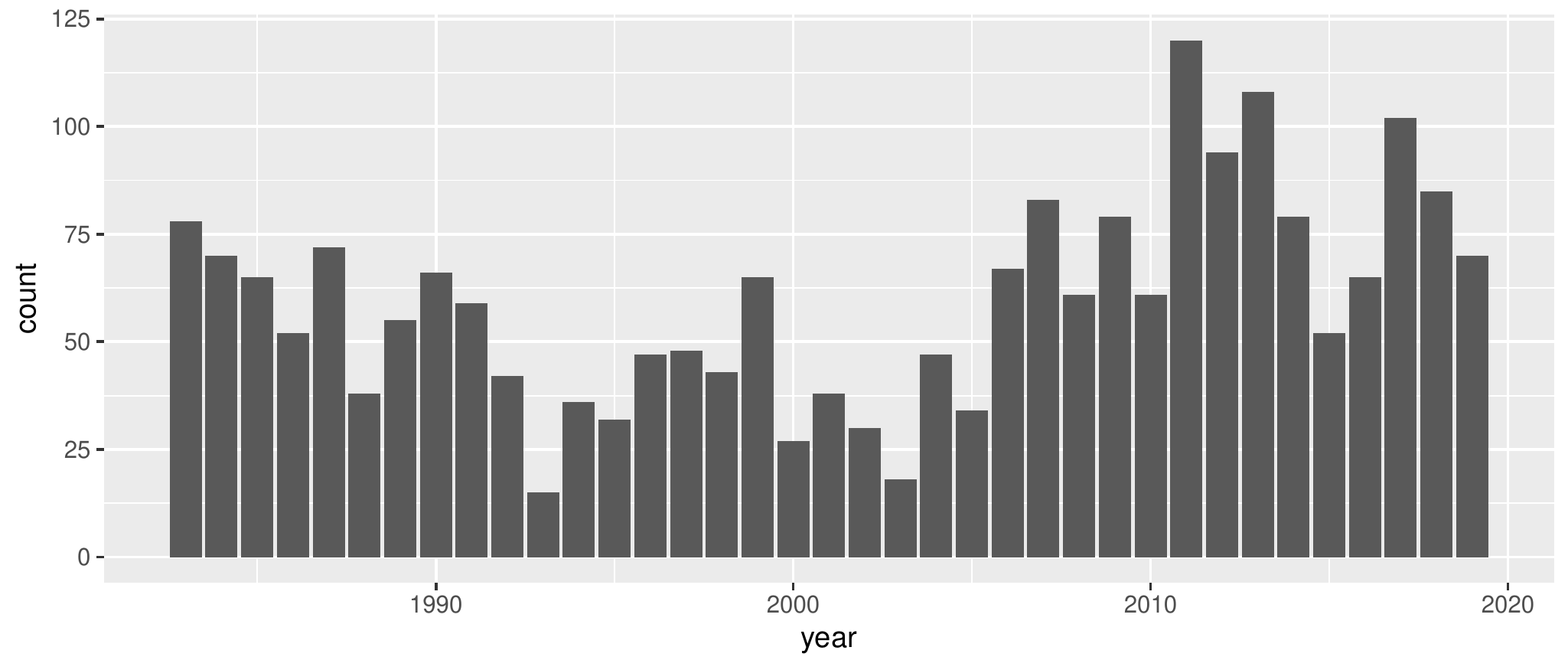}
 \caption{The distribution of Eurasian blackcaps across years.}
 \label{svhyear}
\end{figure}

For the single species models, we chose the species Eurasian Blackcap (\textit{Sylvia atricapilla}), since it is easily classified in terms of \textit{species}, \textit{age} and \textit{sex}. Thus, we have highly reliable data, and it is possible to incorporate the covariates \textit{age} and \textit{sex} in the single species models. The Eurasian Blackcap is quite a common choice of species for phenological analysis, and an interesting modeling approach can be found in \citet{aharon2021limited} for this species, using lqmm with \textit{year} as block factor and fixed effect. The distribution of birds across \textit{age} and \textit{sex} levels are shown in Table \ref{svhtab}, whereas the number of birds recorded each year is illustrated in Figure \ref{svhyear}. Another single species phenological analysis can be found in \citet{hossjer2021on}, with an analysis of Common Redstart {\it Phoenicurus phoenicurus}. In that article we investigate not only changes in location of the arrival distribution, but also changes in scale, skewness and kurtosis, and how covariates affect these.
\begin{table}
\centering
 \begin{tabular}{|c|c|c|}
  \hline & Female & Male \\ \hline
  Juvenile & 567 & 786 \\ \hline
  Adult & 406 & 444 \\ \hline
 \end{tabular}
\caption{Age-sex distribution of the Eurasian blackcap data. There is a larger number of juveniles and males caught, but the imbalance is not particularly extreme.}
\label{svhtab}
\end{table}

\section{Results} \label{results}
First, we look at the single species Eurasian blackcap and compare two ways of estimating a quantile regression model with an empirical quantiles model, in terms of paramater estimates and pointwise confidence interval coverage. Thereafter, we fit models for multiple species, with species as a block factor, and a random intercept and random slope on {\it year} for each species, or as a categorical fixed effect, interacting with {\it year}. These models are scrutinized with special regard to quantile crossing. For all fitted models, {\it year} was centered around 2001.

\subsection{Species specific modelling}\label{sec:single}

\subsubsection{Model setup}
All models fitted to the Eurasian blackcap data set include the same covariates, and (as covered in Section \ref{sec:sspm}) differ in what is used as response variable, and how the objective function is constructed. 
The general specifications of the empirical quantiles (eq) and quantile regression (qr) models are provided in equations \eqref{eqlm} and \eqref{qrmod} respectively. For all models $p=2$, with $\beta_0$ the intercept, $\beta_t$ the slope on year $t$, $\beta_1$ the effect of age $x_1$ and $\beta_2$ the effect of sex $x_2$. The quantile is specified by $\tau$, $\varepsilon$ denotes the error term and there are $n=2\,203$ observations.
Models were fitted for $\tau = 0.01,0.02,\ldots,0.99$. Pointwise bootstrap based approximate confidence intervals were computed for all regression parameters $\hat{\beta}$, for each $\tau$. This was done by resampling pairs $(y,x)$ with replacement until 1000 resampled data sets of the same size as the original had been generated. Each model was fitted to the same resamp\-led data set, to get the most comparable confidence intervals, which were constructed by extracting the $2.5\%$ and $97.5\%$ quantiles from the resulting sample of parameter estimates.

\subsubsection{Estimation methods} \label{sec:estmethod}
For all eq-models we used the maximum likelihood estimates. For the qr-models we used the Frisch-Newton interior point method for optimizing \citep{portnoy1997gaussian} the objective function \eqref{qrest} in order to get our parameter estimates. For the lqms, we used both the gradient based method of \citet{bottai2015gradient}, and Nelder-Mead interior point estimation \citep{nelder1965simplex}. These two optimization approches yielded considerably different estimates. The estimates obtained through Nelder-Mead optimization were very close to the Frisch-Newton estimates for qr, which is to be expected due to the similarity of the two objective functions. However, the gradient based estimates were different from those obtained with the other two optimization methods, for $\hat{\beta}_1(\tau)$ as well as for $\hat{\beta}_2(\tau)$. For this reason, we chose to include the gradient based estimates in Figure \ref{4plots}, to highlight the differences occuring with this optimization method, since the estimates using Nelder-Mead were very close to the qr-estimates.

\subsubsection{Parameter estimates} \label{sec:paramest}
\begin{figure}[!ht]
\centering
\includegraphics[width=\textwidth]{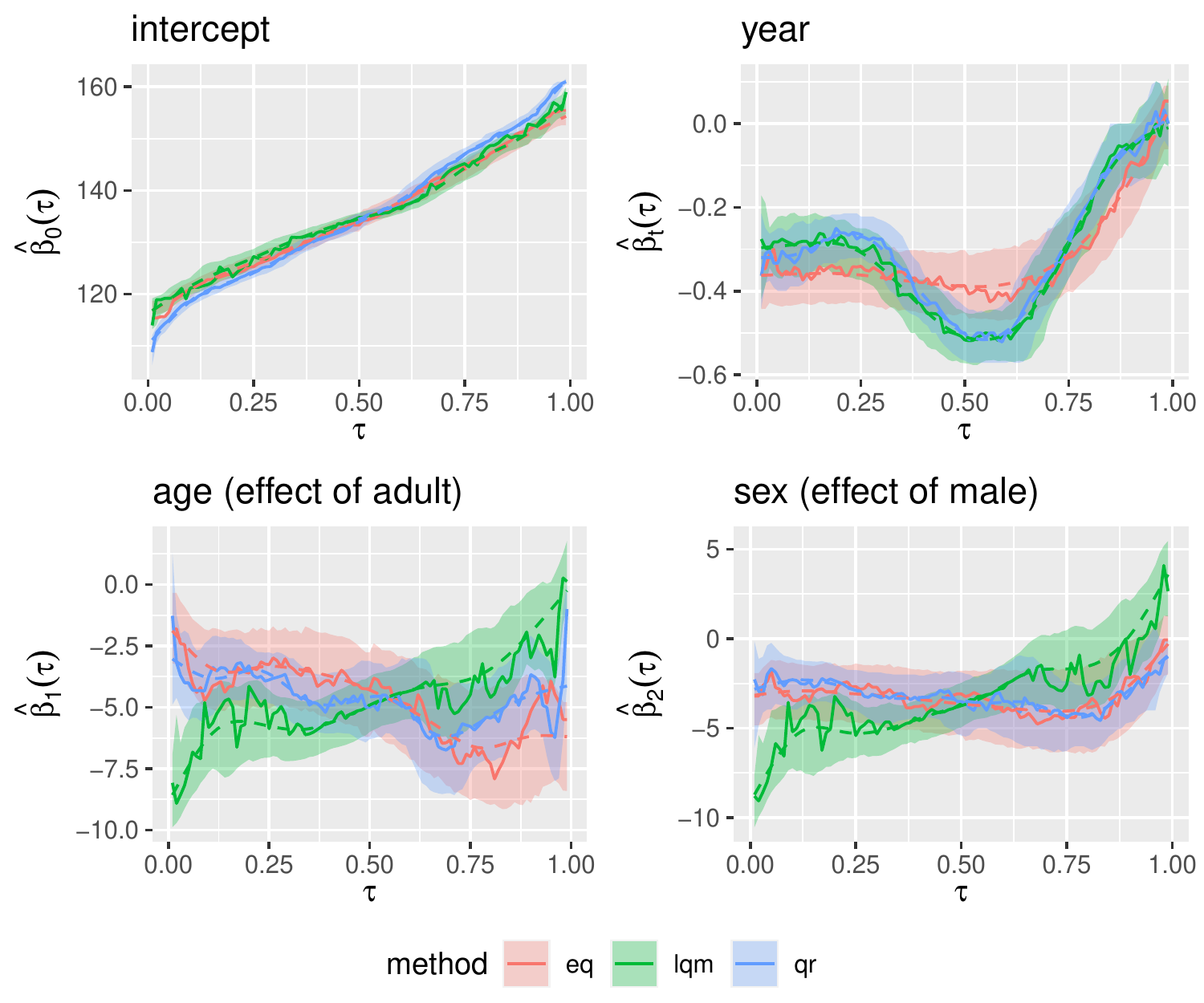}
 \caption{The legend at the top follows the abbreviations already introduced in the article. The solid line consists of the actual parameter estimates for all quantiles, the dashed line is a cubic spline smoothing of all the bootstrap estimates and the shaded band is the interpolated 95\% pointwise confidence interval based on quantiles of the bootstrap distribution. More detailed comments are provided in Section \ref{sec:paramest}.}
 \label{4plots}
\end{figure}

For each method we checked if the intercept estimates were monotonically increasing in $\tau$. It can clearly be seen in Figure \ref{4plots} that $\hat{\beta}^{\text{lqm}}_0(\tau)$ is not monotone. Moreover, there is one instance where $\hat{\beta}^{\text{qr}}_0$ is decreasing with $\tau$, but since the decrease is of magnitude $10^{-14}$, we cannot rule out numerical errors. For eq, the intercept estimates are always monotone in $\tau$. Since {\it year} is centered around 2001, the interpretation of the intercept is the estimated arrival distribution year 2001.
\par\medskip
For the eq model, notice the lacking coverage of the confidence intervals in the tails of the distribution when it comes to the intercept estimate $\hat{\beta}^{\text{eq}}_0$. The estimates of the effect of {\it age} $\hat{\beta}^{\text{eq}}_1$ also approaches the limits of the confidence interval. Although some missing coverage is expected since the confidence band consists of interpolated pointwise approximate 95\% confidence intervals, the large discrepancy between estimates and confidence intervals is not ideal. 
\par\medskip
Further inspection of Figure \ref{4plots} reveals that the intercept estimates are relatively close for all three methods, with $\hat{\beta}^{\text{qr}}_0$ having a bit heavier tails. The estimates $(\hat{\beta}^{\text{qr}}_t,\hat{\beta}^{\text{lqm}}_t)$ of the effect of {\it year} are also close across $\tau$-values, whereas $\hat{\beta}^{\text{eq}}_t$ does not decrease as much in the two middle quartiles. Conversely, the effects of {\it age} $(\hat{\beta}^{\text{qr}}_1, \hat{\beta}^{\text{eq}}_1)$ and {\it sex} $(\hat{\beta}^{\text{qr}}_2, \hat{\beta}^{\text{eq}}_2)$ are similar, whereas $\hat{\beta}^{\text{lqm}}_1$ and $\hat{\beta}^{\text{lqm}}_2$ differ markedly from the former two methods in the lower and upper tertials of the distribution. One can also notice near identical estimates of $\hat{\beta}^{\text{lqm}}_1$ and $\hat{\beta}^{\text{lqm}}_2$ in the lowest quartile. All covariates have a diminishing effect as we approach the right tail of the distribution. This might be due to data containing some local breeding birds, especially towards the end of the sampling window. Since these do not originate from the migration process, they might showcase different patterns.

\begin{table}[ht]
\centering
 \begin{tabular}{r|c|c|c|c}
   Method & intercept & {\it year} & {\it age} & {\it sex} \\
   \hline
   eq & 3.358 & 0.164 & 3.900 & 3.832 \\
   lqm & 3.646 & 0.158 & {\bf 3.207} & 3.514 \\ 
   qr & {\bf 2.939} & {\bf 0.151} & 3.390 & {\bf 3.286}
 \end{tabular}
\caption{The table presents the mean CI width of the single species model, for each combination of method and parameter, when averaged over all values of $\tau$ for which estimation was performed. The shortest mean CI is highlighted with bold font.}
\label{tab:ciw}
\end{table}

\subsubsection{Conclusions} \label{sec:conclusion}
Our overall assessment is that qr produces the most consistent and detailed results, as well as having the overall tightest confidence intervals. As a caution, we do not recommend gradient based optimization when including binary covariates in the quantile regression models. The large deviations in the parameter estimates compared to the bootstrapped estimates for the eq model make us cautions using this method, although it otherwise seems to detect the more or less same signal as the qr model.

\clearpage

\subsection{Multiple species modelling}
For the multiple species models we used the three approaches described in Section \ref{sec:mult}. Since some of the species included lack {\it age} and {\it sex} data, no other fixed effects than {\it year} were included. For the qr-model, {\it species} was a fixed effect interacting with {\it year}. In order to ease the computational burden we only fitted these models to 21 quantiles, namely $(0.01, 0.05, \ldots, 0.95, 0.99)$. For each quantile we predicted random effects $\hat{u}_i^{\text{meq}}$ and $\hat{u}_i^{\text{lqmm}}$ of the meq and lqmm respectively, for each species $i$. 
\par\medskip
For the lqmm we were unable to get the gradient based estimation algorithm \citep{bottai2015gradient} to work on anything but a small subset of the data. Instead we used Nelder-Mead optimization \citep{nelder1965simplex}, implemented in the \texttt{optim}-function in \textsf{R}, which can be called directly from the \texttt{lqmm}-function. As for the single species case, the qr-model was estimated using the Frisch-Newton method \citep{portnoy1997gaussian}. The meq-model was estimated using REML \citep{laird1982random}, described in Section 3.4 of \citet{bates2015fitting}.
\par\medskip
We computed 
\begin{equation}\label{hatbeta}
\begin{split}
\hat{\beta}_i^{\text{meq}}(\tau) &= \hat{\beta}^{\text{meq}}(\tau) + \hat{u}^{\text{meq}}_i(\tau)\\ 
\hat{\beta}_i^{\text{lqmm}}(\tau) &= \hat{\beta}^{\text{lqmm}}(\tau) + \hat{u}^{\text{lqmm}}_i(\tau)
\end{split}
\end{equation}
for $i=1,\ldots,M$ to get species specific intercepts $\hat{\beta}_{i0}(\tau)$ and slopes $\hat{\beta}_{it}(\tau)$ on {\it year} for each model. For the qr-model there are only species specific effects, since {\it species} is a fixed effect interacting with {\it year}. We thus compute $\hat{\beta}_i^{\text{qr}}$ by summing the two coefficients on the right-hand side of \eqref{qrmult}, for the intercept and effect of year respectively. In Figures \ref{icm} and \ref{ym} we have linearly interpolated and plotted the intercepts and effects of {\it year} across all 21 quantiles.

\subsubsection{Intercept}
\begin{figure}[!ht]
 \includegraphics[width=\textwidth]{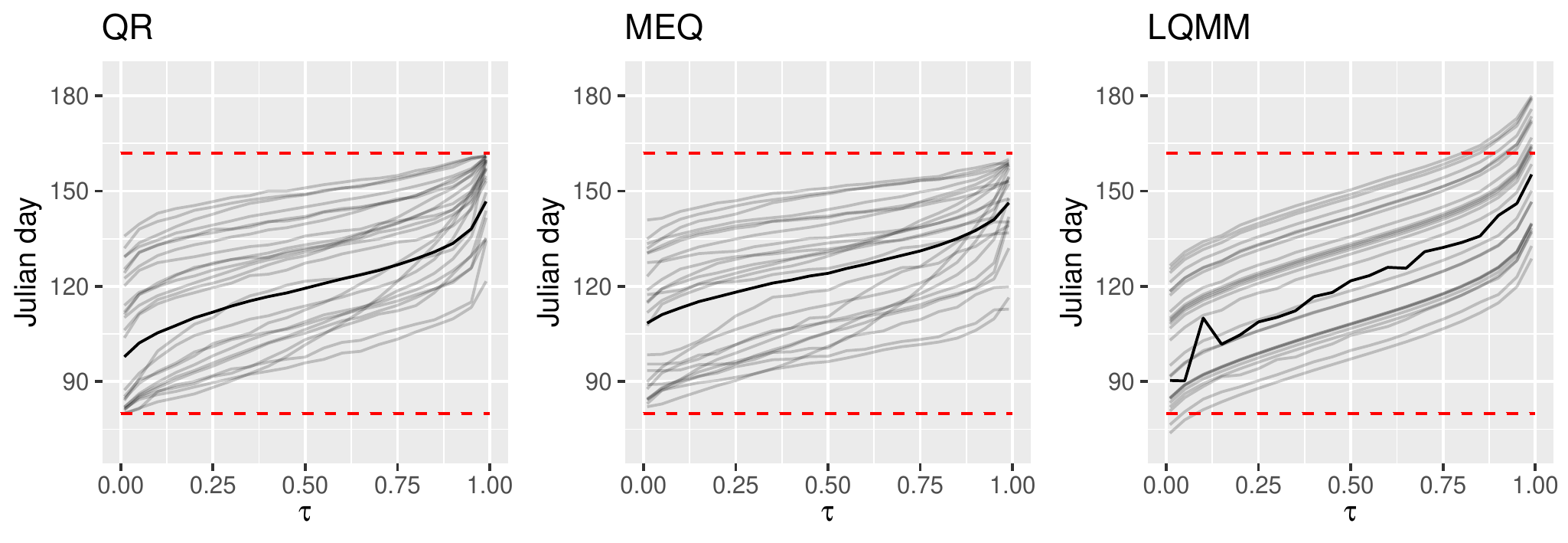}
 \caption{Intercept plots for the multiple species models. The grey lines illustrate the interpolated species specific intercept estimates $\hat{\beta}_{i0}^{\text{qr}}(\tau)$, $\hat{\beta}_{i0}^{\text{meq}}(\tau)$ and $\hat{\beta}_{i0}^{\text{lqmm}}(\tau)$ for $i=1,\ldots,N$. The black line is the overall intercept estimate for the meq and lqmm methods, and for the qr, it is the weighted mean $N^{-1}\sum_{i=1}^M n_i \hat{\beta}_{i0}^{\text{lqmm}}(\tau)$ of the species specific estimates.}
 \label{icm}
\end{figure}

Of the intercept estimates, $\hat{\beta}_{i0}^{\text{qr}}(\tau)$ was monotonically increasing in $\tau$ for all $i$, which implies that the weighted mean also is monotonically increasing. For the lqmm-models $\hat{\beta}_{i0}^{\text{lqmm}}(\tau)$ increased monotonically for all $i$, but $\hat{\beta}^{\text{lqmm}}(\tau)$ did not (visible in Figure \ref{icm} for several quantiles). In view of (\ref{hatbeta}), this indicates that the fixed and random intercept effects of the lqmm-model are difficult to separate for some quantiles $\tau$. For the meq model, there was one species for which $\hat{\beta}_{i0}^{\text{meq}}(0.01) > \hat{\beta}_{i0}^{\text{meq}}(0.05)$, otherwise $\hat{\beta}_{i0}^{\text{meq}}(\tau)$ increased monotonically in $\tau$ for all $i$ and so did $\hat{\beta}^{\text{meq}}(\tau)$.
\par\medskip
The species specific intercept estimates $\hat{\beta}_{i0}^{\text{lqmm}}(\tau)$ of the lqmm stand out in that most of them are close to parallel (since the predicted random intercept effect of each species is virtually independent of $\tau$). This would imply that the only difference between species' arrival distributions is a location shift (i.e. a temporal shift of the distribution). The estimates further stand out in that they exceed the sampling limits at both ends of the sampling window. Some species might indeed be migrating outside the sampling window, but further examination is needed to rule out the possibility that the sampling window limits are exceeded as a consequence of the difficulty of obtaining a flexible, species-specific estimates of the arrival distribution.
\par\medskip
Although $\hat{\beta}_{i0}^{\text{qr}}(\tau)$ and $\hat{\beta}_{i0}^{\text{meq}}(\tau)$ are similar, there is a general difference in that $\hat{\beta}_{i0}^{\text{qr}}(\tau)$ has more steeply sloping tails, especially in the lower quantiles. The overall conclusion of the crossing lines is that the arrival distributions of different species differ in further aspects than location \citep{oja1981location}.

\subsubsection{Year}
\begin{figure}
 \includegraphics[width=\textwidth]{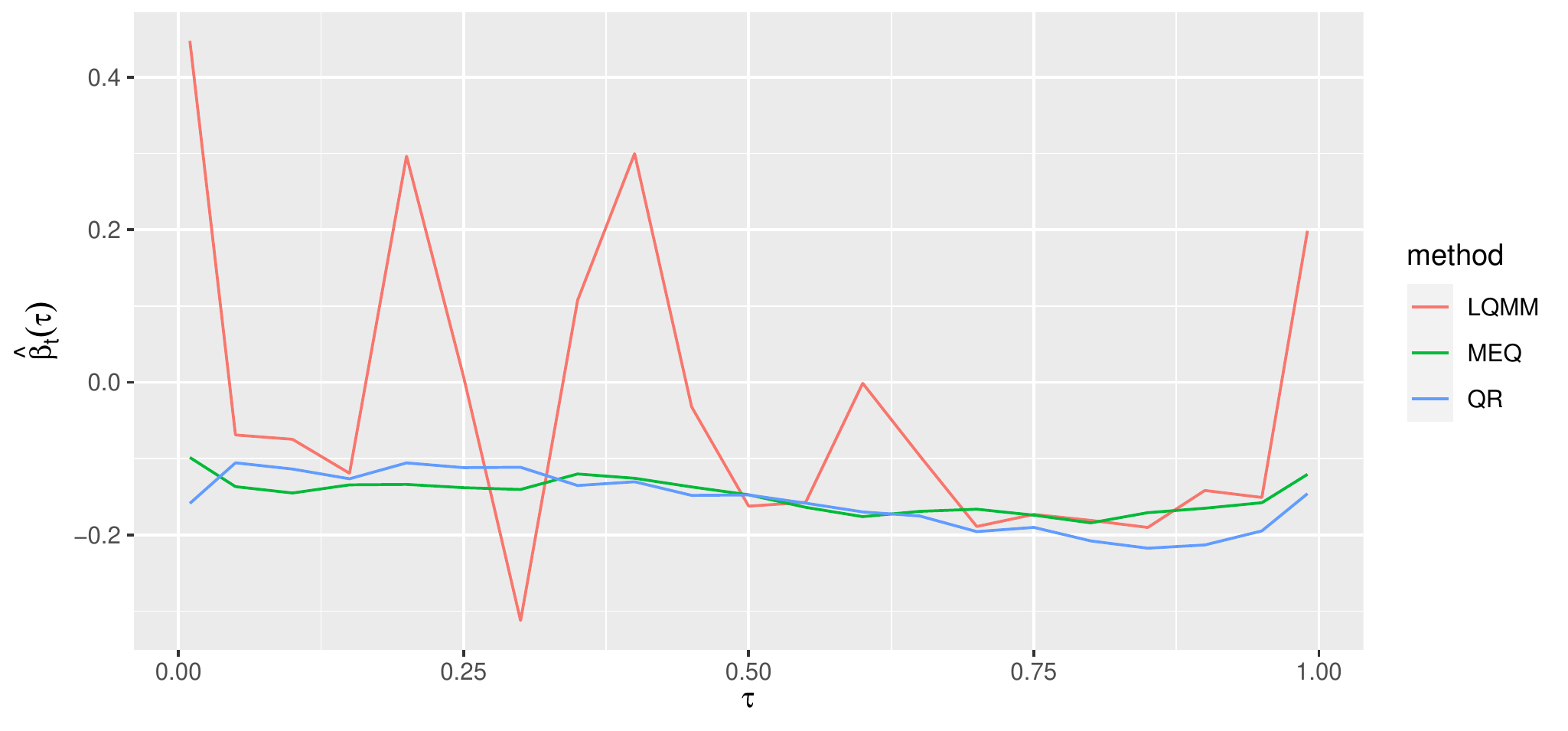}
 \caption{Plots of $\hat{\beta}_t^{\text{lqmm}}(\tau)$, $\hat{\beta}_t^{\text{meq}}(\tau)$ and $\hat{\beta}_t^{\text{qr}}(\tau)$ as a function of $\tau$. Notice the sharp jumps in $\hat{\beta}_t^{\text{lqmm}}(\tau)$, whereas $\hat{\beta}_t^{\text{meq}}(\tau)$ and $\hat{\beta}_t^{\text{qr}}(\tau)$ follow each other closely.}
 \label{ym}
\end{figure}

\begin{figure}
 \includegraphics[width=\textwidth]{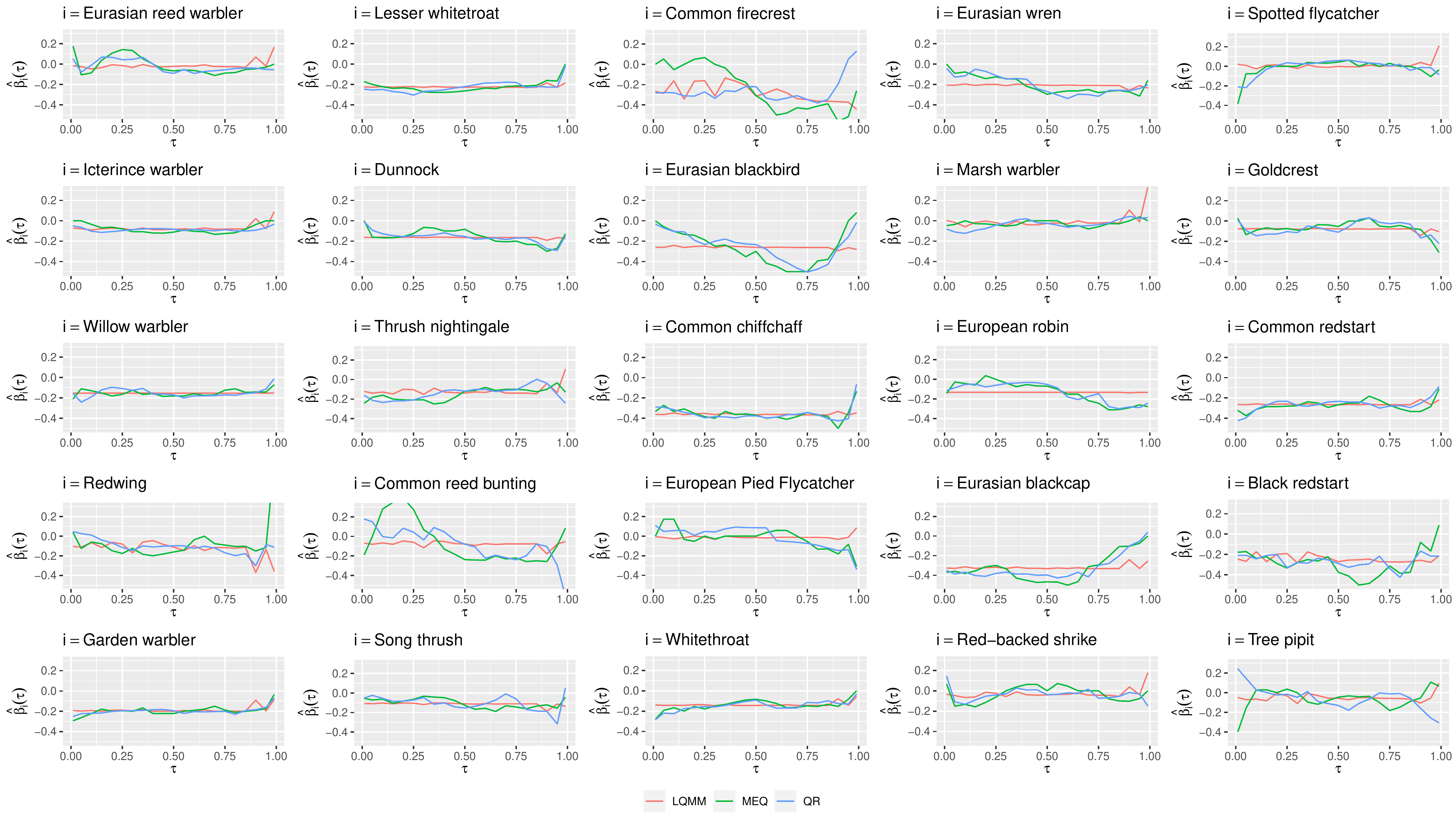}
 \caption{Estimates of $\beta_{it}(\tau)$ for each species of the multiple species model. For a few quantiles and species, $\hat{\beta}_{it}(\tau)$ is located outside the range of the plot. This vertical truncation makes it easier to inspect where $\hat{\beta}_{it}(\tau)$ is not constant in $\tau$. Notice the overall positive correlation between the qr and meq estimates, wheras the lqmm estimates seem more independent of the other methods, and overall are very flat.}
 \label{allsp}
\end{figure}

In Figure \ref{ym} the sharply oscillating behavior of $\hat{\beta}_t^{\text{lqmm}}(\tau)$, as a function of $\tau$, is clearly visible. Combined with the flat lines of $\hat{\beta}_{it}^{\text{lqmm}}(\tau)$ in Figure \ref{allsp}, the conclusion is that $\beta_{it}^{\text{lqmm}}(\tau)$ is possible to estimate, but the separation of $\beta^{\text{lqmm}}$ and $u_i$ is problematic for each quantile $\tau$. Since $\hat{\beta}_t^{\text{lqmm}}(\tau)$ changes value radically with $\tau$, whereas $\beta_{it}^{\text{lqmm}}(\tau)$ is a smooth function of $\tau$ for each species $i$, this implies that $\hat{u}^{\text{lqmm}}_i$ is negatively correlated with $\hat{\beta}^{\text{lqmm}}$, indicating that the fixed and random effects are difficult to separate. 
\par\medskip
If $\beta_{it}(\tau)$ is more or less constant in $\tau$, there is not much need for a quantile regression model. If the location change is the same across all quantiles, an ordinary linear mixed effects model (not using empirical quantiles) would be much more suitable. However, inference based on the eq- and the meq-models indicate that the effect of {\it year} varies with $\tau$ for several species.
\par\medskip
Figure \ref{allsp} shows that for all three models and the majority of species and $\tau$-values, $\hat{\beta}_{it}(\tau)$ are negative. This means that overall, species have advanced their spring arrival. This is consistent with many previously published studies of bird phenology \citep{jonzen2006rapid, tottrup2006patterns, lehikoinen2019phenology, usui2017temporal, aharon2021limited}, here with added details of how this effect changes across the arrival time distribution, both within species and generally across species. As a closing observations we notice that for most species $i$, $\hat{\beta}_{it}^{\text{meq}}(\tau)$ and $\hat{\beta}_{it}^{\text{qr}}(\tau)$ follow each other closely. 

\subsubsection{Conclusion}
The above results indicate that the lqmm framework is unsuitable for a multiple species analysis of bird phenology, at least for our dataset. The overall similarities between the qr and meq-models indicates a common signal being detected in data, and the results seem much more plausible than the lqmm-results. Considering the overall purpose of the phenological analysis at hand, we therefore propose to incorporate {\it species} as a block factor of fixed effect.

\section{Discussion} \label{discussion}

In this article we fitted empirical quantile models and quantile regression models to phenological data of bird migration, in the context of single and multispecies models. Overall the ordinary quantile regression model, without random effects, provided the most stable and reliable results. 
\par\medskip 
For the multispecies analysis based on a quantile regression model with random effects (lqmm), the oscillating behavior of the fixed effects estimates $\hat{\beta}^{\text{lqmm}}(\tau)$, as a function of the quantile $\tau$, is not ideal (as seen in Figures \ref{icm} and \ref{ym}). Further analyses are needed in order to find out whether this stems from an objective function with many local optima (so that the optimization methods could be trapped in one of these), or if the global maximum varies rapidly with $\tau$. Although it is known that the log likelihood objective function of the lqmm is concave \citep{geraci2014linear}, the marginalization of the random effects $u$ is done numerically, using Gaussian quadrature. We cannot rule out that the resulting approximation of the objective function, unless it is very accurate, has local optima. The number of knots used in the Gaussian quadrature can be tuned, and we found that the oscillating behavior of the fixed effect estimates were less pronounced when we set it to 25.
\par\medskip
It would be natural to complement the results of this paper by a simulation study. A good starting point for such a study could be the Negative Binomial approach of \citet[p51]{linden2011using}, which can be employed repeatedly to simulate a multiple species data set. Another approach to simulating phenological data for smoothing can be found in \citet{knudsen2007characterizing}, with potentially influential factors collected in Table 2.
\par\medskip
We chose to not include confidence intervals for the parameters of the multiple species models for two reasons. First, the main focus was to investigate the difference in estimates between the three approaches (qr, meq and lqmm). In this context a goal was to investigate whether lqmm is suitable for phenological analysis of multiple species. Our results indicate that this is not the case. Therefore, and secondly, we did not make the large computational effort to perform the built in block bootstrap approach of the \texttt{lqmm}-package.
\par\medskip
We filtered the data provided by the Falsterbo Bird Observatory so that species with a relatively large number of local breeding birds were omitted. Some birds of the included species might still be local breeding individuals, rather than migratory individuals, and it is likely that this contamination is largest at the end points of the sampling window. In particular, the fact that the estimate $\hat{\beta}_t(\tau)$ of the fixed effect of year approches 0 as the quantile $\tau$ approaches 1, might be caused by non-migratory individuals. Ideally, one would filter data for birds that probably are not migrating, or at least assign weights to birds depending on how likely it is that they are migrating. Such a weighting scheme could employ covariates such as fat score \citep{rogers1991evaluation} or data on repeated captures of birds within a season.
\par\medskip
It would also be ideal to include birds that are recaptured between seasons. These recaptures generally constitute a very small fraction of the number of first captures, and probably they have a propensity for being local breeders. In spite of this, birds recaptured between seasons might give deeper insight into the variation in arrival date within individual birds, as as an instance of repeated measurements.
\par\medskip
At the beginning of the sampling window some species might already have begun their migration, and at the end of the sampling window some species might still be arriving. This results in incomplete measurements for the arrival distribution of these species. It is important to account for such incomplete data when presenting measures of location, scale, skewness and kurtosis for the arrival distributions. Such $L$-functionals measures have been used by \cite{hossjer2021on}, for quantile regression models, in the context of single species phenological analysis. These methods are also applicable to censored and truncated data.
\par\medskip
The equal weighting of birds in this paper aims at creating a complete picture of the passage of migratory birds. A natural alternative would be to weight species equally, and thus focus the study on patterns in the behaviors of various species. In some studies it might be prefable to focus on the general behavior of species rather than birds. For the multiple species models, such a shift is easily accomodated by adjusting the species specific weights.

\section*{Acknowledgements}
The authors would like to thank the Falsterbo Bird Observatory for providing the data.

\appendix

\section{Efficient matrix inversion for $\hat{u}$} \label{AppMatrix}
For all fitted models in this paper, $q=0$. Thus, this appendix will cover random effects estimation for the specific case that each group (species) has a random intercept and a random slope. Omitting the random slope will make the computations less involved, and adding further random slopes will complicate it. Several grouping factors are not supported by \texttt{lqmm}.
\par\medskip 
Recall that species $i$ has $n_i$ observations, and now denote the number of observed birds of a particular species year $t$ as $n_t$, $t=1,\ldots,T$. We will consider one species at a time, and therefore omit the species index $i$ in the notation, unless it appears in $n_i = \sum_{t=1}^T n_t$. The arrival time of bird $j$ year $t$ is then
\begin{equation}
 y_{tj} = x_{tj}\beta(\tau) + u_{1} + u_{2}t + \varepsilon_{tj}(\tau),
\end{equation}
where $x_{tj}=(1,t,x_{j1},\ldots,x_{jp})$ includes the covariates of bird $j$ from year $t$. The arrival time of all birds of the particular species can be written in matrix form as
\begin{equation}
 Y = X \beta(\tau) + Zu + \varepsilon(\tau)
\end{equation}
where $Y$ and $\varepsilon(\tau)$ are column vectors that contain the arrival times $y_{tj}$ and the error terms $\varepsilon_{tj}(\tau)$ of all birds, stacked yearwise on top of each other,   
\begin{align}
 Z = \begin{pmatrix}
         & 1_{n_1} \\
        1_{n_i} & \vdots \\
         & 1_{n_T}T
       \end{pmatrix}, \quad 
u_i = \begin{pmatrix}
        u_{i1} \\
        u_{i2}
       \end{pmatrix}, \quad
 \Psi(\tau) = \begin{pmatrix}
      \psi_{11} & \psi_{12} \\
      \psi_{21} & \psi_{22} 
     \end{pmatrix}
\end{align}
and $1_{n_\cdot}$ denotes a column vector with ones of length $n_\cdot$.
We introduce
\begin{equation}
 B = Z \Psi(\tau) Z^\top = \begin{pmatrix}
      B_{11} & \dots & B_{1T} \\
      \vdots & \ddots & \vdots \\
      B_{T1} & \dots & B_{TT}
     \end{pmatrix}
\end{equation}
where
\begin{align}
 B_{st} &= \lh \psi_{11} + \psi_{12}\lp s+t \rp + \psi_{22}st \rh \lp1_{n_s} \otimes 1_{n_t}\rp \\
 &:= b_{st} \lp1_{n_s} \otimes 1_{n_t}\rp.
\end{align}
This means that each block $B_{st}$ in $B$ is a matrix of dimension $n_s \times n_t$ with value $b_{st}$ everywhere. The repeated occurence of the same values within blocks is the structure we will make use of to make the matrix inversion more efficient, by finding the inverse to a reduced version of $\Sigma$ and then rebuilding it using the block structure. We first postulate that $\Sigma^{-1}$ should have the same type of block structure as $\Sigma$, i.e.
\begin{equation} \label{sigmainverse}
 \Sigma^{-1} = C + \delta_\varepsilon I_{n_i}.
\end{equation}
Here, $C$ is a matrix with the same structure as $B$, but with $b_{st}$ replaced with $c_{st}$, and $\delta_\varepsilon > 0$ a scalar. If we first find $\delta_\varepsilon$ and all $c_{st}$, we can then make use of \eqref{sigmainverse} and compute $\Sigma^{-1}$ from these values and $\lb n_t\rb_{t=1}^T$. From the equality
\begin{equation} \label{matrixequation}
 \lp B + \sigma_\varepsilon I_{n_i} \rp \lp C + \delta_\varepsilon I_{n_i}  \rp =
 BC + \delta_\varepsilon B + \sigma_\varepsilon C + \sigma_\varepsilon\delta_\varepsilon I_{n_i} = I_{n_i} 
\end{equation}
we introduce
\begin{equation} \label{Amat}
 BC := A = \begin{pmatrix}
      A_{11} & \dots & A_{1T} \\
      \vdots & \ddots & \vdots \\
      A_{T1} & \dots & A_{TT}
     \end{pmatrix}
\end{equation}
and consequently, $A_{st} = a_{st} \lp1_{n_s} \otimes 1_{n_t}\rp$, where
\begin{equation} \label{ast}
 a_{st} = \sum_{r=1}^T n_r b_{sr}c_{rt}
\end{equation}
which can be seen by performing the matrix multiplication in \eqref{Amat}. Our goal is to find a matrix of reduced dimension to invert, so to this end we introduce the matrices 
\begin{equation} \label{checkmatrices}
 \check{A} = \begin{pmatrix}
      a_{11} & \dots & a_{1T} \\
      \vdots & \ddots & \vdots \\
      a_{T1} & \dots & a_{TT}
     \end{pmatrix} \quad
\check{B} = \begin{pmatrix}
      b_{11} & \dots & b_{1T} \\
      \vdots & \ddots & \vdots \\
      b_{T1} & \dots & b_{TT}
     \end{pmatrix} \quad
\check{C} = \begin{pmatrix}
      c_{11} & \dots & c_{1T} \\
      \vdots & \ddots & \vdots \\
      c_{T1} & \dots & c_{TT}
     \end{pmatrix}
\end{equation}
and the diagonal matrix $\check{N}$ with diagonal elements contained in the vector $\text{tr}(\check{N}) = \lp n_1,\ldots,n_T\rp$.  In conjunction with \eqref{ast} this gives the equality 
\begin{equation}
 \check{A} = \check{B}\check{N}\check{C}.
\end{equation}
Applying the dimension reduction ${\cal M}^{n_i \times n_i} \mapsto {\cal M}^{T \times T}$ to the RHS of \eqref{matrixequation} we get
\begin{equation} \label{matrixequationreduced}
  \check{B}\check{N}\check{C} + \delta_\varepsilon \check{B} + \sigma_\varepsilon \check{C} + \sigma_\varepsilon\delta_\varepsilon I_{T} = I_{T}.
\end{equation}
A solution to \eqref{matrixequationreduced} is
\begin{equation} \label{sol}
 \begin{cases}
  \sigma_\varepsilon\delta_\varepsilon = 1, \\
  \check{B}\check{N}\check{C} + \delta_\varepsilon \check{B} + \sigma_\varepsilon \check{C} = 0_T \otimes 0_T,
 \end{cases}
\end{equation}
where the upper equation of \eqref{sol} gives $\delta_\varepsilon=\sigma_\varepsilon^{-1}$, whereas the lower equation of \eqref{sol} yields 
\begin{align}
 \check{C} &= -\delta_\varepsilon \lp\check{B}\check{N} + \sigma_\varepsilon \rp^{-1} \check{B}.
\end{align}
With $\delta_\varepsilon$ and $\check{C}$ determined, we plug them into \eqref{sigmainverse} to find $\Sigma^{-1}$, which is then inserted into \eqref{ueblp}.
\par\medskip
However, if $n_i$ is large, the memory requirements of \eqref{ueblp} might still lead to a crash on some computers, and thus we adapted \eqref{ueblp} to compute $\hat{u}^{(\tau)}$ sequentially. We divide $\Sigma^{-1}$ into $T$ submatrices $\lh \Sigma^{-1}_1 | \ldots | \Sigma^{-1}_T \rh$ where each submatrix consists of adjacent columns of $\Sigma^{-1}$ of dimension $n_i \times n_t$ for $t=1,\ldots,T$. We then compute for each $t$ the $2 \times n_t$ matrices
\begin{equation}
 W_t = \hat{\Psi}^{(\tau)} Z^\top \hat{\Sigma}^{-1}_t
\end{equation}
and combine them into the $2\times n_i$ matrix $W = \lh W_1|\ldots|W_T\rh$. As a final step we then compute 
\begin{equation}
\hat{u}^{(\tau)} = W \lb y - X\hat{\beta}^{(\tau)} - \hat{\mathbb{E}}\lh \varepsilon^{(\tau)} \rh \rb.
\end{equation}
The whole procedure is still quite memory intensive, but we manage to run it on a computer with 32GB of RAM with the most numerous species having 38 738 observations. For species with a few thousand observations, the sequential procedure is not needed for computing \eqref{ueblp}, but the dimension reduced approach to inverting $\Sigma$ is recommended, as it speeds up computation considerably.
\par\medskip
The only added requirement on data for our particular implementation (cf.\ the supplementary material), is that birds are ordered by species, and ordered chronologically within species.

\newpage
\bibliography{qrrefs}
\end{document}